\documentclass[11pt]{article}
\usepackage[centertags]{amsmath}
\usepackage{graphicx,indentfirst,amsmath,amsfonts,amssymb,amsthm,newlfont}

\usepackage[latin1]{inputenc}

\def\al{\alpha}

\def\var{\varphi}

\def\l1{{\lambda}_1}

\newcommand{\f}{\frac}

\def\x1{{\xi }_{xx}}
\def\x2{{\xi }_{yy}}
\def\x3{{\xi }_{xy}}

\def\e1{{\eta }_{xx}}
\def\e2{{\eta }_{yy}}
\def\e3{{\eta }_{xy}}

\def\kd{\partial}

\newcommand{\ds}{\displaystyle }

\newcommand{\beqn}{\begin{eqnarray*}}
\newcommand{\eeqn}{\end{eqnarray*}}
\newcommand{\beqnn}{\begin{eqnarray}}
\newcommand{\eeqnn}{\end{eqnarray}}
\newcommand{\p}{\partial}
\newcommand{\bb}{\begin{equation}}
\newcommand{\ee}{\end{equation}}
\newcommand{\ba}{\begin{array}}
\newcommand{\ea}{\end{array}}

\begin{document}
\pagenumbering{arabic}
\title{\huge \bf Group Analysis of the Novikov Equation}
\author{\rm \large Yuri Bozhkov$^1$, Igor Leite Freire$^{2}$ and Nail H. Ibragimov$^{3,4}$\\ \\
\it $^1$Instituto de Matem\'atica,
Estat\'\i stica e \\ \it Computa\c c\~ao Cient\'\i fica - IMECC \\
\it Universidade Estadual de Campinas - UNICAMP \\ \it Rua S\'ergio Buarque de Holanda, 651 \\  \it $13083$-$859$ - Campinas - SP, Brasil
\\ \rm E-mail: bozhkov@ime.unicamp.br \\ \\
\it $^2$Centro de Matemática, Computação e Cognição\\ \it Universidade Federal do ABC - UFABC\\ \it
Rua Santa Adélia, 166, Bairro Bangu\\ \it $09.210-170$ - Santo André, SP - Brasil\\
\rm E-mails: igor.freire@ufabc.edu.br and igor.leite.freire@gmail.com\\ \\
\it $^3$Laboratory ``Group analysis of mathematical models
in natural and engineering sciences''\\
\it Ufa State Aviation Technical University\\
\it $450 000$ Ufa, Russia\\ \\
\it $^4$ Department of Mathematics and Science\\
\it Blekinge Institute of Technology\\
\it $SE-371 79$ Karlskrona, Sweden\\
\rm E-mails: nailhib@gmail.com and nib@bth.se} 
\date{\ }
\maketitle \vspace{.5cm}
\begin{abstract}
We find the Lie point symmetries of the Novikov equation and
demonstrate that it is strictly self-adjoint. Using the
self-adjointness and the recent technique for constructing conserved
vectors associated with symmetries of differential equations, we
find the conservation law corresponding to the dilations symmetry
and show that other symmetries do not provide nontrivial
conservation laws. Then we investigat the  invariant solutions.
\end{abstract}
\vskip 0.5cm
\begin{center}
{2010 AMS Mathematics Classification numbers:\vspace{0.2cm}\\
76M60, 58J70, 35A30, 70G65\vspace{0.2cm} }
\end{center}
\pagenumbering{arabic}
\newpage

\section{Introduction}

\

Since the celebrated Camassa-Holm equation
\bb\label{ch}
u_{t}+2\kappa u_{x}-u_{xxt}+3uu_{x}=2u_{x}u_{xx}+uu_{xxx}
\ee
was derived in \cite{cholm}, intense research dealing with integrable non-evolutionary partial differential equations of the type
\bb\label{gen}
u_{t}-u_{txx}=F(u,u_{x},u_{xx},u_{xxx},\cdots)
\ee
has been carried out. Such equations are used in modelling shallow water waves.

During sometime the Camassa-Holm equation was the only known example of integrable equation of the type (\ref{gen}) having solutions as a superposition of multipeakons, that is, peaked soliton solutions with discontinuous derivatives at the peaks. In \cite{deg} it was proved that an equation obtained previously by Degasperis and Procesi also had such property.

In a recent communication \cite{nov} a classification of integrable equations of the type (\ref{gen}) with quadratic and cubic nonlinearities was carried out. In the same paper the new partial differential equation
\bb\label{u01}
u_t - u_{txx} + 4u^2 u_x - 3 u u_x u_{xx} - u^2 u_{xxx}=0
\ee
containing cubic nonlinearities was discovered by V. S. Novikov and is named after him. Equation (\ref{u01}) is considered as a type of generalization of (\ref{ch}). Since then equation (\ref{u01}) has been the subject of intense research, see \cite{him,hone1,hone2,ni1,ni2,w,wu} and references therein.

Namely, in \cite{him} the authors study the Cauchy problem for the Novikov equation, well-posedness and dependence on initial data in Sobolev spaces. Local well-posedness in the Besov spaces for the Cauchy problem for equation (\ref{u01}) was proved in \cite{ni2}. In \cite{ni1} a global existence result and conditions on the initial data were considered. Existence and uniqueness of global weak solution to Novikov with initial data under some conditions was proved in \cite{wu}. A generalization of (\ref{u01}) with dissipative term was considered in \cite{w}. Peakon solutions were studied in \cite{hone1,hone2}.

The purpose of this paper is to study the Novikov equation with the methods of the modern group analysis. First we find the Lie point symmetries of (\ref{u01}) (section 2). In section 3 we prove that the Novikov equation is strictly self-adjoint. Then we establish the conservation laws corresponding to the already found symmetries using the new conservation theorem \cite{i4} providing conservation laws for any  differential equation, inclusive those which do not possess a Lagrangian as well as those of odd order. This is done in section 4. We obtain and discuss some invariant solutions in section 5.

The presentation of the results is very condensed and summarized. The technical details are ommited in order not to increase the volume of this paper, which, otherwise should triplicate.

\

\section{Lie point symmetries}

\

Application of the classical technique for calculating the
symmetries of differential equations shows that the infinitesimal
point symmetries  of the Novikov equation (\ref{u01}) span the
five-dimensional Lie algebra with the following basis:


\[ X_1= \frac{\kd}{\kd t},\;\;\; X_2=\frac{\kd}{\kd x},\;\;\; X_3 = e^{2x} \frac{\kd}{\kd x} + e^{2x} u \frac{\kd}{\kd u},\]
\[ X_4 = e^{-2x} \frac{\kd}{\kd x} - e^{-2x} u \frac{\kd}{\kd u},\;\;\; X_5 = -2t \frac{\kd}{\kd t} + u \frac{\kd}{\kd u}.\]

It is easy to see that the discrete transformation $x\mapsto-x$ maps the Lie point symmetry generator $X_{3}$ into $X_{4}$.

\ 

\section{Self-adjointness}

\

The Novikov equation (\ref{u01}) does not have a classical Lagrangian. Therefore, following \cite{i4,i6}, we introduce the formal Lagrangian ${\cal L}=vF$, where $v=v(t,x)$ is the nonlocal variable and $F$ denotes the
left-hand side of (\ref{u01}), namely

\bb\label{sa1} \mathcal{L} =v [u_t - u_{txx} + 4u^2 u_x - 3 u u_x u_{xx} - u^2 u_{xxx} ].\ee

Then applying the Euler operator to ${\cal L}$ we obtain that the adjoint equation to (\ref{u01}) reads:

\bb\label{sa2} F^* = -v_t + v_{txx} - 4 u^2 v_x +3 u v_x u_{xx} - 3 v u_xu_{xx} + 3 u u_x v_{xx} +u^2v_{xxx}.\ee

Hence $F^*|_{v=u}= -F$. Therefore the Novikov equation (\ref{u01}) is strictly self-adjoint.

Of course, the latter fact can be obtained by using the concept of
nonlinear self-adjointness \cite{i7}
and the substitution
$v=\var(t,x,u)$ which imply $v=\al u$, where $\al $ is a constant.
This fact reflects another commom property of the Camassa-Holm and
Novikov equations, since both are strictly self-adjoint. The strict
self-adjointness of equation (\ref{ch}) was proved in \cite{i5}.

\

\section{Conservation laws}

\

Let
$$X=\xi^{0}\f{\p}{\p t}+\xi^{1}\f{\p}{\p x}+\eta\f{\p}{\p u}$$
be an infinitesimal generator of a Lie point symmetry admitted by equation (\ref{u01}).
According to the conservation theorem proved by Ibragimov in \cite{i4}, it is possible to find nonlocal conservation laws for equation (\ref{u01}) of the form $D_{t}C^{0}+D_{x}C^{1}=0$ , where
\bb\label{3.2.0}
\ba{lcl}
C^{i}&=&W\,
\ds{\left[\f{\p{\cal L}}{\p u_{i}}-D_{j}\left(\f{\p{\cal L}}{\p u_{ij}}\right)+D_{j}D_{k}\f{\p{\cal L}}{\p u_{ijk}}\right]}\\
\\
&&\ds{+D_{j}(W)\,\left[\f{\p{\cal L}}{\p u_{ij}}-D_{k}\left(\f{\p{\cal L}}{\p u_{ijk}}\right)\right]+D_{j}D_{k}(W)\,\f{\p{\cal L}}{\p u_{ijk}}},
\ea
\ee
$i,j,k=0,1,\,x^{0}=t,\,x^{1}=x,\,W=\eta-\xi^{i}u_{i}$, $D_i$ is the total derivative operator with respect to $x^i$, the formal Lagrangian (\ref{sa1}) is written
in the symmetric form
\bb\label{cl1} \mathcal{L} =v [u_t - \frac{1}{3}(u_{txx}+u_{xtx}+u_{xxt}) + 4u^2 u_x - 3 u u_x u_{xx} - u^2 u_{xxx} ]\ee
and we have used the Einstein summation convention. Moreover, one can reduce the conserved vector $C=(C^0,C^1)$ by
applying the simplifying procedure described on pages 50-51 of \cite{i7}. We shall use this approach to establish
the conservation laws for the third order evolution equation (\ref{u01}).

Since the Novikov equation is strictly self-adjoint we can eliminate the nonlocal variable $v$ into the components $C^{i}$ given
in (\ref{3.2.0}). Then we can establish a local conservation law for equation (\ref{u01}).

\

{\bf Theorem.} {\it $(i)$ The Lie point symmetry $X_5$ of the Novikov equation provides the conserved vector $C=(C^0,C^1)$ with components}
\bb\label{cl4}\ba{lll} C^0 &=& u^2 + u_x^2,\\
& & \\
C^1 &=&2 u^4 - 2 u^3 u_{xx}-2 u u_{tx}.\ea\ee

{\it $(ii)$ For the symmetries $X_1,X_2,X_3$ and $X_4$ the conserved vector $C=(C^0,C^1)$ is the null vector.}

\

{\bf Proof.} \rm (i) From (\ref{3.2.0}) the component $C^0$ is determined by
\bb\label{cl2}
C^{0}=W\,
\ds{\left[\f{\p{\cal L}}{\p u_{t}}+D_{x}^2\f{\p{\cal L}}{\p u_{txx}}\right]-(D_x W) .D_x\f{\p{\cal L}}{\p u_{txx}}+(D_x^2W). \f{\p{\cal L}}{\p u_{txx}}},
\ee
where $W=u+2tu_t$ for the symmetry $X_5$. Substituting $W$ and $\cal L$ from (\ref{cl1}) into (\ref{cl2}) we obtain
\bb\label{cl3}
C^0|_{v=u} = (u+2tu_t) (u-\f{1}{3}u_{xx}) + (u_x+2tu_{tx})\f{1}{3}u_x -\f{1}{3}u(u_{xx}+2tu_{txx}).
\ee
Further we use the identities
\[uu_{xx}=D_x(u u_x) - u_x^2,\,\,\,\,u_xu_{tx}=D_x(u_t u_x) - u_t u_{xx}, \]
\[ D_x(u u_{tx}-u_tu_x) = u_xu_{tx}-u_t u_{xx},\,\,\,\,\,uu_{txx} = D_x(u u_{tx}-u_tu_x) +u_tu_{xx},\]
as well as the equation (\ref{u01}) from which we express $u_t$ to get from (\ref{cl3}), after cancelation of some terms, that
\bb\label{cl5} C^0|_{v=u} = u^2 + u_x^2 + D_x[ \f{4}{3} tuu_{tx} - \f{2}{3}tu_tu_x -\f{2}{3}u u_x +2tu^3u_{xx}-2t u^4].\ee
Hence the first component can be reduced to that given in (\ref{cl4}).

Now we transfer the third term in (\ref{cl5}) to the second component $C^1$ according to \cite{i7}, pp. 50-51. In this way we arrive at
\[ \ba{lll} \tilde{C}^1|_{v=u}& =& C^1 +D_t[ \f{4}{3} tuu_{tx} - \f{2}{3}tu_tu_x -\f{2}{3}u u_x +2tu^3u_{xx}-2t u^4]\\
& & \\
&=& (u+2tu_t)[4u^3 -3u^2 u_{xx}-\f{2}{3} u_{tx}] + \f{1}{3}u_t (u_x+2tu_{tx})+\f{1}{3}u_x(3u_t+2tu_{tt})\\
& & \\
& & - \f{2}{3}u(3u_{tx}+2tu_{ttx})- (u_{xx}+2tu_{txx})u^3 \\
& & \\
& & +D_t[ \f{4}{3} tuu_{tx} - \f{2}{3}tu_tu_x -\f{2}{3}u u_x +2tu^3u_{xx}-2t u^4].
\ea
\]
Hence after differentiating with respect to $t$ in the last term above and simplifying we obtain that $C^1$ has the form given in (\ref{cl4}).

(ii) In a similar way by a starightforward and tedious work one can see that the Lie point symmetries $X_1,X_2,X_3$ and $X_4$ give trivial conservation laws. This completes the proof of the Theorem.

\

We would like to observe that the verification of the conservation law $D_{t}C^{0}+D_{x}C^{1}=0$ where the components of the vector $C=(C^0,C^1)$ are determined by (\ref{cl4}) is very simple. Indeed, the following identity holds:
\bb\label{cl7}D_{t}(u^2 + u_x^2) +D_{x}(2 u^4 - 2 u^3 u_{xx}-2 u u_{tx})=
2u(u_t - u_{txx} + 4u^2 u_x - 3 u u_x u_{xx} - u^2 u_{xxx}). \ee

The identity (\ref{cl7}) allows us to recover the proof of Lemma 2.6 in \cite{w} where an ad hoc procedure was used without any reference to symmetries and conservation laws while we have established (\ref{cl4}) by a systematic application of the new conservation theorem \cite{i4}. In fact, if $u(0,x)=u_0(x)$, then we can easily obtain from (\ref{cl7}) that the solutions of (\ref{u01})) satisfy the identity
\[\int_{\mathbb{R}}  (u^2 + u_x^2)dx=\int_{\mathbb{R}}  (u^2_0 + u_{0x}^2)dx \]
assuming the functions $u$ and $u_0$ belong to certain functions spaces such that the integrals exist.

\

\section{Invariant solutions}

\

In this section we obtain the invariant classical solutions corresponding to the Lie point symmetries found in Section 2. For this purpose we shall apply the well known procedure described in \cite{i1,ol} without presenting the straightforward details.

\begin{itemize}
\item The invariant solution for the symmetry $X_{1}$ (translation in $t$) is simply a function of $x$ only. We then substitute $u=u(x)$ into (\ref{u01}), multiply by $u$ and obtain the ODE
\[ 4u^3 u' - 3 u^2 u' u'' - u^3 u'''=0,\]
where $'=d/dx$, which is equaivalent to
\[ (u^4-u^3u'')'=0.\]
This equation can be integrated at once obtaining
\bb\label{aux} u^3u''=u^4 + A,\ee
where $A$ is an integration constant. The solutions to (\ref{aux}) are given by 
\bb\label{is9}
\ba{lcl}
u_{1}^{\pm}(x)&=& \ds{\pm\frac{1}{2} e^{-(x+c_{2})} \sqrt{4 A+e^{4( x+ c_{2})}-2 e^{2 (x+ c_{2})} c_{1}+c_{1}^2}},\\
\\
u_{2}^{\pm}(x)&=& \ds{ \pm\f{1}{2}\sqrt{4Ae^{2( x+ c_{2})} +e^{-2( x+ c_{2})} -2c_{1}+c_{1}^2e^{2(x+c_{2})}}}.
\ea
\ee

Whenever $A=0$ equation (\ref{aux}) is linear and its general solution is $u(x)=c_{1}e^{x}+c_{2}e^{-x}$, where $c_{1}$ and $c_{2}$ are arbitrary constants.

\item The invariant solutions for the symmetry $X_{2}$ (translation in $x$) are just constants.

\item The invariant solutions corresponding to the symmetry $ X_3 = e^{2x} \frac{\kd}{\kd x} + e^{2x} u \frac{\kd}{\kd u}$ are of the form
\bb\label{is1}
u(t,x)=\var (t)e^x
\ee
where $\var =\var (t)$ is any one time differentiable function of $t$.

\item Similarly, the invariant solutions corresponding to the symmetry $ X_4 = e^{-2x} \frac{\kd}{\kd x} - e^{-2x} u \frac{\kd}{\kd u}$ are of the form
\bb\label{is2}
u(t,x)=\var (t)e^{-x}
\ee
where $\var =\var (t)$ is any one time differentiable function of $t$. This result reflects a discrete symmetry correspondence between $X_3$ and $X_4$.

\item Regarding the Lie point symmetry $X_5 = 2t \frac{\kd}{\kd t} - u \frac{\kd}{\kd u}$ the invariant solutions are of the form
\[ u(t,x)=\frac{1}{\sqrt{t}}\psi(x)\]
where $\psi =\psi (x)$ satisfies the folllowing equation
\bb\label{is12} 4{\psi}^2 {\psi}' - 3 {\psi} {\psi}' {\psi}'' - {\psi}^2 {\psi}'''+\f{1}{2}{\psi}''-\f{1}{2}{\psi}=0.\ee 
Clearly $\psi(x)=e^{\pm x}$ is a solution of (\ref{is12}) and thus two invariant solutions are given by 
\[ u_{\pm}(t,x) =\frac{1}{\sqrt{t}}e^{\pm x}. \]

\item Looking for travelling wave solutions $u=\phi(x-ct)$, that is, solutions invariant with respect to
$X_2-cX_1=\frac{\kd}{\kd x}-c\frac{\kd}{\kd t}$, where $c$ is a constant to be determined, one can obtain the ODE:
\bb\label{is5} 4{\phi}^2 {\phi}' - 3 {\phi} {\phi}' {\phi}'' - {\phi}^2 {\phi}'''+c{\phi}''-c{\phi}'=0.\ee

where $z=x-ct$ and $'=d/dz$. An obvious solution to (\ref{is5}) is $\phi=e^{z}.$ Then the function
$$u(x,t)=e^{x-ct}$$
is a travelling wave solution to the Novikov equation.
\end{itemize}

\noindent
\textbf{Remark.} Using a general separation of variables
\bb\label{sep} u(t,x)=\var(t)\psi(x)\ee
one obtains that
\bb\label{is7} {\var}'=k{\var}^3\ee
and
\bb\label{is8} 4{\psi}^2 {\psi}' - 3 {\psi} {\psi}' {\psi}'' - {\psi}^2 {\psi}'''= k({\psi}''-{\psi}), \ee
for some constant $k$ provided ${\psi}''-{\psi}\neq0$. (Otherwise $u(t,x)={\var}_1(t) e^x+{\var}_2(t) 
e^{-x}$, a case already seen.) Compare also with the invariant solutions under $X_1$ ($k=0$) and $X_5$ ($k=-1/2$) discussed above. The equation (\ref{is7}) can be immediately solved. However the equations 
(\ref{is5}) and (\ref{is8}) and their solutions remain to be investigated thoroughly. 

\section{Conclusion}

In this paper the recent new equation (\ref{u01}) derived by V. S. Novikov
\cite{nov} was studied from the point of view of Lie point
symmetries theory. From the developments due to one of us (Nail Ibragimov
 \cite{i4,i6,i7})
we proved that only one symmetry, namely the dilation symmetry,
provides a nontrivial local conservation law.

An infinite number of invariant solutions to the Novikov equation were obtained, see equations (\ref{is1}) and (\ref{is2}).

\subsection*{Acknowledgements}

\

The authors would like to thank FAPESP, S\~ao Paulo, Brasil, and BTH, Sweden, for the support giving Nail H. Ibragimov the opportunity to visit IMECC-UNICAMP, where this work was initiated. Yuri Bozhkov would also like to thank FAPESP and CNPq, Brasil, for partial financial support. Igor Leite Freire is thankful to IMECC-UNICAMP for gracious hospitality and UFABC for the financial support. N. H. Ibragimov's work is partially supported by the Government of Russian Federation through Resolution No. 220, Agreement No. 11.G34.31.0042.


\begin{thebibliography}{99}





\bibitem{cholm} R. Camassa and D.D. Holm, An integrable shallow water equation with peaked solitons, Phys. Rev. Lett, 71, (1993) 1661--1664.

\bibitem{deg} A. Degasperis, D. D. Holm and A. N. W. Hone, A new integrable equation with peakon solutions, Theor. Math. Phys., 133 (2002) 1461--1472.

\bibitem{him} A. A. Himonas and C. Holliman, The Cauchy problem for the Novikov equation, Nonlinearity, 25 (2012) 449--479.

\bibitem{hone1} A. N. W. Hone and J. P. Wang, Integrable peakon equations with cubic nonlinearity, J. Phys. A: Math. Theor. 41 (2008) 372002, 10 pp.

\bibitem{hone2} A. N. W. Hone, H. Lundmark, and J. Szmigielski, Explicit multipeakon solutions of Novikov's cubically nonlinear integrable Camassa-Holm type equation, Dyn. Partial Differ. Equ. 6 (2009), 253--289.


\bibitem{i1} N. H. Ibragimov, Transformation groups applied to mathematical physics,  Translated from the Russian  Mathematics and its Applications (Soviet  Series), D. Reidel Publishing Co., Dordrecht, (1985).



\bibitem{i4} N. H. Ibragimov, A new conservation theorem, J. Math. Anal. Appl. 333 (2007) 311--328.

\bibitem{i5} N. H. Ibragimov, R. S. Khamitova, A. Valenti, Self-adjointness of a generalized Camassa-Holm equation, Appl. Math. Comp. 218 (2011) 2579--2583.

 \bibitem{i6}  N. H. Ibragimov, Nonlinear self-adjointness and conservation laws, J. Phys. A: Math. Theor. 44 (2011) 432002, 8 pp.

\bibitem{i7} N. H. Ibragimov, Nonlinear self-adjointness in constructing conservation laws, Archives of ALGA 7/8 (2010-2011)  1--90.
See also \textit{arXiv}:1109.1728v1[math-ph], 2011, pp. 1-104.

\bibitem{ni1} Z Jiang and L. Ni, Blow-up phenomenon for the integrable Novikov equation, J. Math. Anal. Appl., 385 (2012) 551--558.

\bibitem{ni2} L. Ni and Y. Zhou, Well-posedness and persistence properties for the Novikov equation, J. Diff. Equ., 250 (2011) 3002--3021.

\bibitem{nov} V. S. Novikov, Generalizations of the Camassa-Holm equation, J. Phys. A: Math. Theor, 42 (2009) 342002, 14pp.

\bibitem{ol}  P. J. Olver, Applications of Lie groups to differential equations, Springer, New York, (1986).

\bibitem{w} Wei Yan, Yongsheng Li, Yimin Zhang, Global existence and blow-up phenomena for the weakly dissipative Novikov equation, Nonlinear Anal. Series A: Theory and Applications 75 (2012) 2464--2473, DOI: 10.1016/j.na.2011.10.044

\bibitem{wu} X. Wu and Z. Yin, Global weak solutions for the Novikov equation, J. Phys. A: Math. Theor, 44 (2011) 055202, 17pp.



\end{thebibliography}
\end{document}